# A Spatio-Temporal Graph Convolutional Network for Gesture Recognition from High-Density Electromyography

Wenjuan Zhong
*Southern University of Science and Technology*
*Department of Biomedicial Engineering*
Shenzhen, China
zhongwj@mail.sustech.edu.cn

Yuyang Zhang
*Southern University of Science and Technology*
*Department of Biomedicial Engineering*
Shenzhen, China
12010722@mail.sustech.edu.cn

Peiwen Fu
*Southern University of Science and Technology*
*Department of Biomedicial Engineering*
Shenzhen, China
12013036@mail.sustech.edu.cn

Wenxuan Xiong
*Southern University of Science and Technology*
*Department of Biomedicial Engineering*
Shenzhen, China
12112550@mail.sustech.edu.cn

Mingming Zhang
*Southern University of Science and Technology*
*Department of Biomedicial Engineering*
Shenzhen, China
zhangmm@sustech.edu.cn

*Abstract*— **Accurate hand gesture prediction is crucial for effective upper-limb prosthetic limbs control. As the high flexibility and multiple degrees of freedom exhibited by human hands, there has been a growing interest in integrating deep networks with high-density surface electromyography (HD-sEMG) grids to enhance gesture recognition capabilities. However, many existing methods fall short in fully exploit the specific spatial topology and temporal dependencies present in HD-sEMG data. Additionally, these studies are often limited number of gestures and lack generality. Hence, this study introduces a novel gesture recognition method, named STGCN-GR, which leverages spatio-temporal graph convolution networks for HD-sEMG-based human-machine interfaces. Firstly, we construct muscle networks based on functional connectivity between channels, creating a graph representation of HD-sEMG recordings. Subsequently, a temporal convolution module is applied to capture the temporal dependences in the HD-sEMG series and a spatial graph convolution module is employed to effectively learn the intrinsic spatial topology information among distinct HD-sEMG channels. We evaluate our proposed model on a public HD-sEMG dataset comprising a substantial number of gestures (i.e., 65). Our results demonstrate the remarkable capability of the STGCN-GR method, achieving an impressive accuracy of 91.07% in predicting gestures, which surpasses state-of-the-art deep learning methods applied to the same dataset.**

## I. INTRODUCTION

Prosthetic Human-Machine Interface (HMI) systems hold great promise in significantly improving the lives of individuals facing amputated limbs or neuromuscular disorders [1]. Particularly for upper-limb functions, human hands exhibit remarkable skill and precision with multiple degrees of freedom. Due to the hands' high flexibility and diversity tasks they perform, achieving an intuitive and seamless control of prostheses poses considerable challenges. To overcome these challenges, most prosthetic HMI systems are designed using gesture recognition algorithms that rely on biological signals recorded from human body [1, 2].

Surface electromyography (sEMG) has been extensively utilized in the literature to facilitate myoelectric control of bionic limbs, enabling the noninvasive peripheral interfacing of human motor intention with robotic actions [1, 3]. The conventional approach involves extracting temporal and spectral features from sEMG signals and feeding them to classic machine learning models such as Support Vector Machines (SVMs) or Linear Discriminant Analysis (LDA) [4]. Decision trees have also been extensively utilized to explore both linear and non-linear relationships encountered in sEMG-based HMIs [5]. While these traditional approaches have shown promise, they inherently face limitations as they require hand-craft features, which may not fully capture the complexities present in sEMG data.

With the exponential growth of data and significant strides in computing power, deep learning technology has rapidly advanced, yielding remarkable achievements in the domain of sEMG-based HMIs [6-8]. Convolutional Neural Networks (CNNs) have proven to be highly effective and scalable tools, capable of discerning patterns in diverse tasks without the need for manual feature selection. For instance, Geng et al. [7] contributed the high-density sEMG (HD-sEMG) dataset CapgMyo, comprising 168 channels. Treating the sampled data at each time point as a 168-pixel image, they harnessed a 2D CNN-based model, employing majority voting to successfully classify gestures. Similarly, Hu et al. [8] proposed an attention-based CNN-RNN model for decode gestures from HD-sEMG images. However, while CNNs excel in handling grid-like inputs within Euclidean space, such as natural images, they may not be as suitable for processing data in non-Euclidean spaces, such as HD-sEMG images. Movement execution necessitates the coordinated activation of numerous muscle groups, and the muscle activation

The work was supported in part by the National Natural Science Foundation of China under Grant 62273173, in part by the Natural Science Foundation of Shenzhen under Grant JCYJ20210324104203010, in part by Shenzhen Key Laboratory of Smart Healthcare Engineering under Grant ZDSYS20200811144003009, in part by the National Key Research and Development Program of China under Grant 2022YFF1202500 and Grant 2022YFF1202502, in part by the Research Foundation of Guangdong Province under Grant 2020ZDZX3001 and Grant 2019ZT08Y191, and in part by the Southern University of Science and Technology.

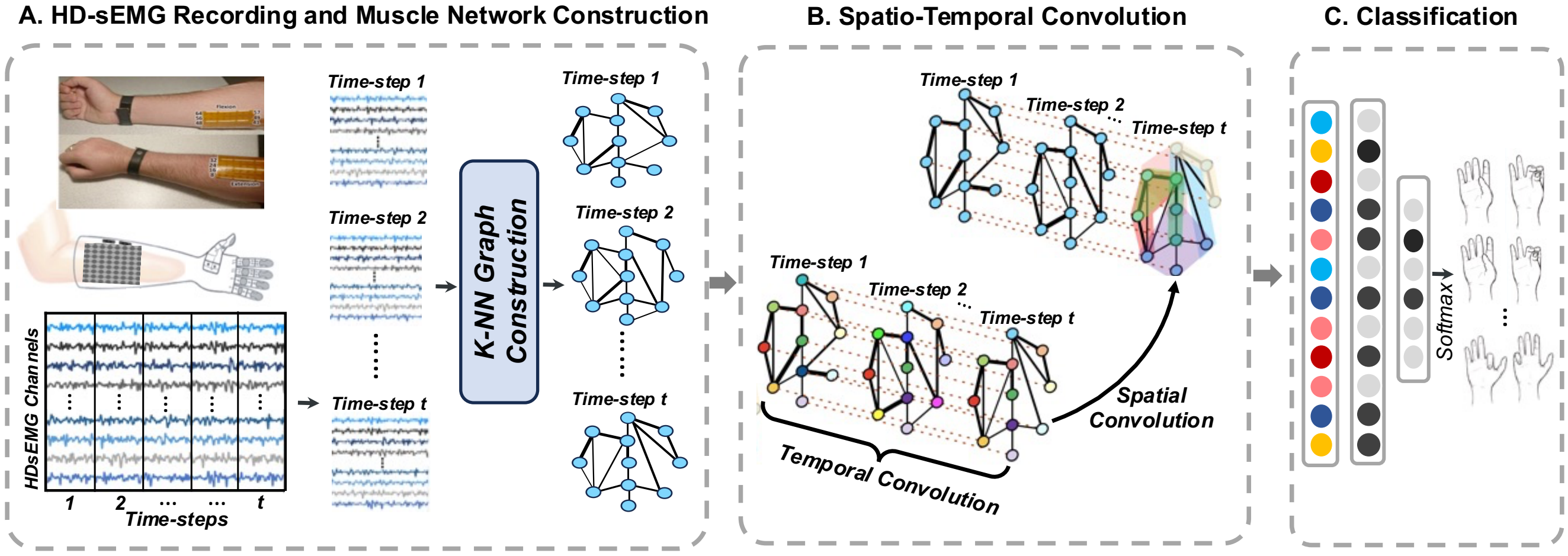

Fig. 1. The framework of the proposed STGCN-GR method. A. Formarm HD-sEMG signals are recorded, and muscle networks are constructed based on the k-NN strategy. B. The spatio-temporal convolution module efficiently captures both topological relations and temporal dependencies within the muscle networks, producing a spatio-temporal feature as the output. C. Fully connected layers are utilized for gesture classification.

between certain distant locations can exhibit high correlations. Consequently, the geometric distance metrics commonly used in Euclidean space may not adequately capture the functional distance between muscle groups.

To address the challenges posed by non-Euclidean data types, such as graphs, several geometric deep learning methods have emerged, including graph convolutional networks (GCNs) [9] and graph attention networks (GATs) [10]. These approaches have shown promise in diverse applications [11], and in particular, GCN has been utilized in brain decoding, where it models the brain as a graph, with regions of interest (ROIs) as nodes and their functional connectivity as edges [12, 13]. Notably, GCNs have also been employed for the first time in HD-sEMG based gesture recognition, demonstrating an impressive 91.25% accuracy in classifying 65 gestures from a shuffled dataset [14]. However, two limitations deserve consideration. Firstly, the strategy of muscle graph construction, which connects only nodes within a heuristic distance, may not fully capture the intricacies of muscle coordination during movements. While GCNs provide valuable insights into graph-based data representation, adopting a heuristic distance-based approach might not effectively account for the functional connectivity strengths between muscle groups. Employing more refined methodologies, such as considering the underlying functional connectivity strengths [15, 16], holds the potential to significantly enhance the accuracy and generalizability of gesture recognition systems. Secondly, this study focuses on processing the spatial correlation between different nodes, overlooking the crucial temporal dynamics inherent in HD-sEMG signals. Recognizing and incorporating the temporal dynamics into the model could lead to further improvements in gesture recognition performance [2, 8], enabling a more comprehensive understanding of how gestures unfold over time. In order to incorporate temporal dependencies into GCN, spatio-temporal graph convolution networks (STGCN) have been developed and applied to traffic forecasting [17]. So far, its application to gesture recognition has not been explored.

In this study, we introduce a Spatio-Temporal Graph Convolution based Gesture Recognition (STGCN-GR) method to further explore the benefits of GCNs in HD-sEMG based gesture recognition. The main contributions of our research are mainly summarized as following:

(1) STGCN-GR integrates a graph convolution module and temporal convolution network, allowing it to effectively capture both topological relations and temporal dependencies within muscle networks constructed from HD-sEMG signals.

(2) Extensive experiments are conducted on a dataset of 65 gestures recorded with HD-sEMG [18]. The results demonstrate that STGCN-GR surpasses state-of-the-art deep learning methods applied to the same dataset, highlighting its superior performance.

(3) Notably, the proposed STGCN-GR successfully decodes the substantial number of gestures (i.e., 65) with sliding window sizes of 250 ms, exceeding the real-time implementation requirement of 300 ms.

## II. Methos

### A. Gesture Recgnition on Muscle Graphs

Decoding motion intentions is a typical classification problem, i.e. predicting the most likely motion intentions, such as hand gestures or gait events. In our research, we focus on studying the HD-sEMG-based motion-decoding problem, which involves classifying hand gestures using the HD-sEMG signals. The framework for this study is thoughtfully illustrated in Fig. 1, providing a clear visual representation of our approach and methodology.

The recorded HD-sEMG time series data denoted as $\mathcal{D} = \{(\boldsymbol{x}_i, y_i)\}_{i=1}^{n}$. Each $\boldsymbol{x}_i \in \mathbb{R}^{N \times T}$ represents task-related HD-sEMG time series data recorded from $N$ electrodes over $T$ time samples. The corresponding $y_i$ represents the task label from label set $Y$, and $n$ is the total number of gestures in the dataset. The objective is to acquire an optimal graph network, denoted as $P(\bullet)$, using the training dataset. This network will facilitate the prediction of the task label $\hat{y}_i$ for each HD-sEMG data $\boldsymbol{x}_i$ present in the testing dataset, as indicated in equation (1).

$$\hat{y}_i = \arg \max_{y_i \in Y} P(y_i \mid \boldsymbol{x}_i) \qquad (1)$$

In our study, we introduce the concept of a muscle network represented as a graph $\mathcal{G}$. The recorded HD-sEMG signals $\boldsymbol{x}_i$ exhibit interdependence and are interconnected through pairwise connections in a graph structure. To construct the graph $\mathcal{G}$, we utilized HD-sEMG to form muscle network graph $\mathcal{G}_t = (\mathcal{V}_t, \mathcal{E}, \boldsymbol{W})$ at the $t$-th time-step. The nodes $\mathcal{V}_t$ are the observations from $N$ electrode in a muscle network; The edges $\mathcal{E}$ are the connectedness between nodes. $\boldsymbol{A} \in \mathbb{R}^{n\times n}$ denotes the weighted adjacency matrix of $\mathcal{G}_t$. The weighted adjacency matrix is obtained by Pearson correlation of HD-sEMG recordings $\boldsymbol{x}_i$ between $N$ electrodes, thus we obtain the fully connected muscle networks.

Subsequently, the k-nearest neighbors (k-NN) strategy is employed to establish connections for each electrode with the top $k$ correlated electrodes using undirected graph edges. In simpler terms, we trim the weighted adjacent matrix $\boldsymbol{A}$ to create a sparse adjacent matrix $\boldsymbol{W}$, following the process described in equation (2). This pruning step helps construct a more efficient and focused graph representation of the interdependences between HD-sEMG electrodes.

$$\boldsymbol{W}_{i,j} = \begin{cases} \boldsymbol{A}_{i,j}, & \text{if } \boldsymbol{A}_{i,j} \geq k^{th}\text{-top } (\boldsymbol{A}_i), \forall \boldsymbol{A}_{i,j} \in \boldsymbol{A}_i \\ 0, & \text{otherwise.} \end{cases} \quad (2)$$

## B. *Spatio-Temporal Graph Convolutional Network for Gesture Recognition*

The primary focus of our proposed method is to efficiently capture both topological relations and temporal dependencies present in muscle networks created from HD-sEMG signals. To achieve this, we present a novel approach called Spatio-Temporal Graph Convolutional Network for Gesture Recognition (STGCN-GR), which incorporates a spatial graph convolution network and a temporal convolution network, inspired by the temporal information forecasting graph networks [17]. The STGCN-GR takes multi-channel HD-sEMG series as input and generates a spatio-temporal feature as output. In the following sections, we provide a detailed description of each module in our method.

(1) Spatial Graph Convolution Module

GCNs update vertices in a graph by gathering information from neighboring vertices (masking with same color in Fig.1) through spatial edges. This mechanism allows GCNs to effectively capture topological structure information, making them highly suitable for processing non-Euclidean data. Interestingly, HD-sEMG signals can be naturally interpreted as graph-structured data, where each electrode corresponds to a node, and functional connections serve as edges. As a result, employing GCNs for HD-sEMG-based gesture recognition can capitalize on the inherent graph-like nature of the data, leading to improved performance in analyzing and interpreting muscle activity patterns.

The spatial graph convolution operator $*_\mathcal{G}$ can be defined as the operation of multiplying an input vector $\boldsymbol{x}$ with a filter $\Theta$.

$$\Theta *_\mathcal{G} \boldsymbol{x} = \Theta(\boldsymbol{L})\boldsymbol{x} = \Theta(\boldsymbol{U\Lambda U}^T)\boldsymbol{x} = \boldsymbol{U}\Theta(\boldsymbol{\Lambda})\boldsymbol{U}^T\boldsymbol{x}, \quad (3)$$

where $\boldsymbol{U}$ and $\boldsymbol{\Lambda}$ are the eigenvector matrix and the diagonal matrix of eigenvalues of the normalized graph Laplacian $\boldsymbol{L} \in \mathbb{R}^{N\times N}$, respectively. We have $\boldsymbol{L} = \boldsymbol{U\Lambda U}^T$.

The graph Laplacian $\boldsymbol{L}$ is from the transformation of the adjacency matrix $\boldsymbol{W}$ with $\boldsymbol{L} = \boldsymbol{I}_N - \boldsymbol{D}^{-\frac{1}{2}}\boldsymbol{W}\boldsymbol{D}^{-\frac{1}{2}}$, where $\boldsymbol{D}$ is the diagonal degree matrix with $\boldsymbol{D}_{i,i} = \sum_j \boldsymbol{W}_{i,j}$, and $\boldsymbol{I}_N$ is an identity matrix.

To enhance the filter localization and reduce the number of parameters, a widely used approach is to employ Chebyshev polynomials $T_k(\bullet)$ to approximate the filter $\Theta$. As a result, the spatial graph convolution can be expressed as a linear function of Chebyshev polynomial $T_k(\tilde{\boldsymbol{L}})$,

$$\Theta *_\mathcal{G} \boldsymbol{x} = \Theta(\boldsymbol{L})\boldsymbol{x} \approx \sum_{k=0}^{K} \theta_k T_k(\tilde{\boldsymbol{L}})\boldsymbol{x}, \quad (4)$$

where $\tilde{\boldsymbol{L}} = \frac{2L}{\lambda_{max}} - \boldsymbol{I}_N$, and $\lambda_{max}$ denotes the largest eigenvalue of $\boldsymbol{L}$.

By introducing the first-order approximation, we set that $\lambda_{max} = 2$ and $\theta = \theta_0 = -\theta_1$. Substituting $\widetilde{\boldsymbol{W}} = \boldsymbol{W} + \boldsymbol{I}_N$, and $\widetilde{\boldsymbol{D}}_{i,i} = \sum_j \widetilde{\boldsymbol{W}}_{i,j}$ into (4), we can approximate the spatial graph convolution as

$$\begin{aligned} \Theta *_\mathcal{G} \boldsymbol{x} &\approx \theta_0 \boldsymbol{x} - \theta_1 \left(\boldsymbol{D}^{-\frac{1}{2}}\boldsymbol{W}\boldsymbol{D}^{-\frac{1}{2}}\right)\boldsymbol{x} \\ &\approx \theta\left(\boldsymbol{I}_N + \boldsymbol{D}^{-\frac{1}{2}}\boldsymbol{W}\boldsymbol{D}^{-\frac{1}{2}}\right)\boldsymbol{x} \\ &\approx \theta\left(\widetilde{\boldsymbol{D}}^{-\frac{1}{2}}\widetilde{\boldsymbol{W}}\widetilde{\boldsymbol{D}}^{-\frac{1}{2}}\right)\boldsymbol{x}. \end{aligned} \quad (5)$$

(2) Temporal Convolution Module

The temporal convolution module is designed to incorporate a 1D convolution with a width-$K_t$ kernel, which is followed by a gated linear unit (GLU) to introduce non-linearity. When processing each node in graph $\mathcal{G}$, the temporal convolution explores $K_t$ neighboring elements of input without padding, resulting in a reduction of sequence length by $K_t - 1$ each time. The input of the temporal convolution for each node can be represented as a sequence of length $l$ with $c_i$ channels, denoted as $\boldsymbol{u} \in \mathbb{R}^{l\times c_i}$. The convolution kernel, denoted as $\Gamma \in \mathbb{R}^{K_t\times c_i \times 2c_o}$ maps $\boldsymbol{u}$ to a single output element $[\boldsymbol{PQ}] \in \mathbb{R}^{(l-K_t+1)\times(2c_o)}$ (where $[\boldsymbol{PQ}]$ is split equally into two parts with channels $c_o$, $c_o$ denotes output channels). Consequently, the temporal gated convolution module can be represented as follows:

$$\Gamma *_\mathcal{T} \boldsymbol{u} = \boldsymbol{P} \odot \sigma(\boldsymbol{Q}) \in \mathbb{R}^{(l-K_t+1)\times c_o}, \quad (6)$$

where $*_\mathcal{T}$ denotes temporal convolution; $\boldsymbol{P}$ and $\boldsymbol{Q}$ are the input of gates in GLU, respectively; $\odot$ denotes the element-wise Hadamard product; $\sigma(\boldsymbol{Q})$ denotes the sigmoid gate which determines the importance of input $\boldsymbol{P}$.

# III. EXPERIMENTS

## A. *Dataset and Preprocessing*

To develop an efficient prosthetic control interface that can support various activities of daily living (ADLs) beyond basic hand functions, this study relies on a high-quality HD-sEMG dataset, recently published in Nature's scientific data [18]. This database consists of 65 isometric hand gestures, each exhibiting different degrees of freedom (DoF). These movements encompass 16 1-DoF finger and wrist gestures, 41 2-DoF compound gestures involving both fingers and the wrist, and eight multi-DoF gestures like grasping, pointing, and pinching [18].

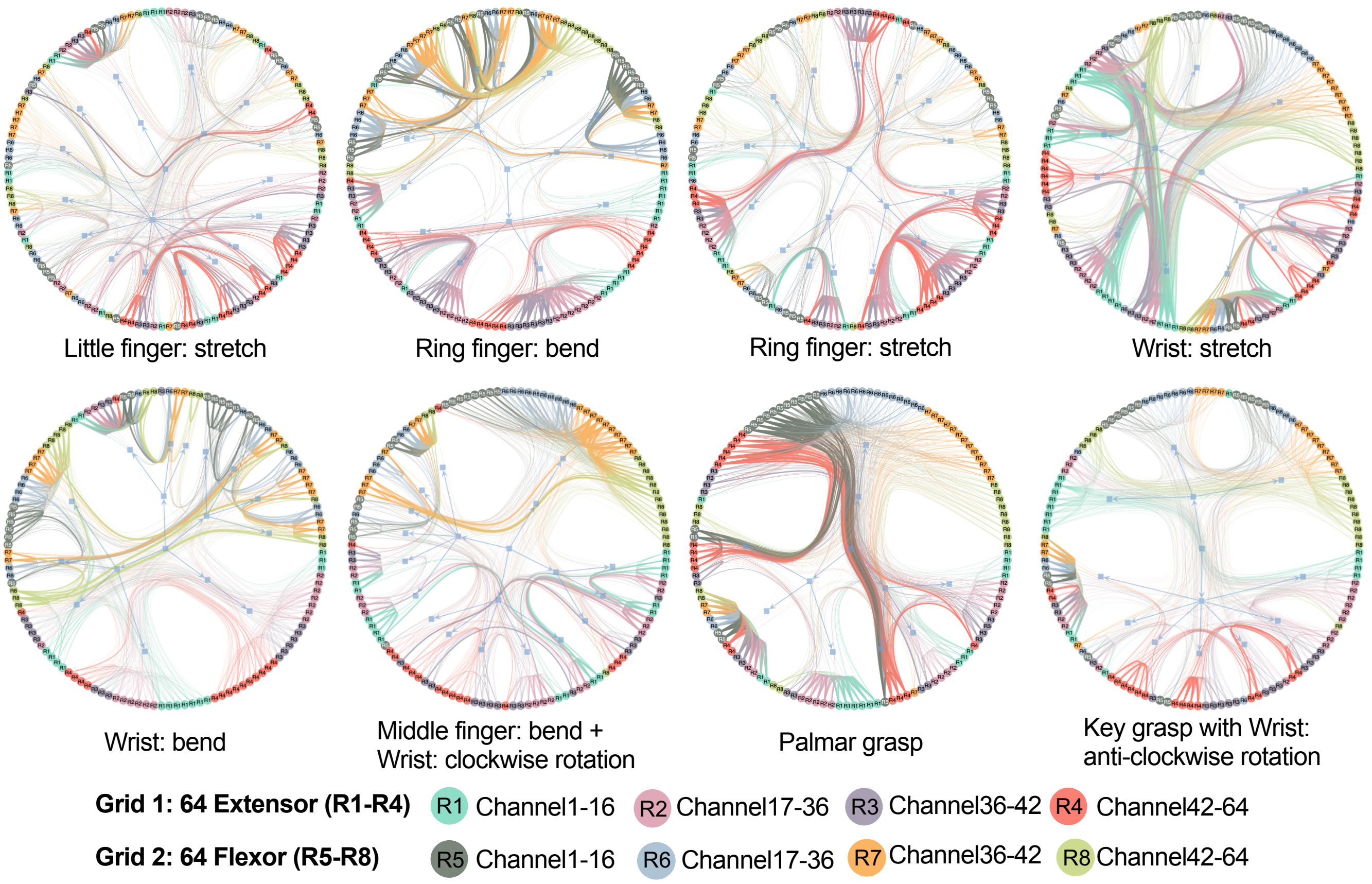

Fig. 2. Graph presentation of full-connected muscle networks for eight gestures of one subject (#20). Thicker edges represent a stronger correlation between nodes. Each graph consists of 128 nodes (64 forearm extensor and 64 forearm flexor), and they are divided into eight regions according to the node locations.

The dataset was collected from 20 healthy participants (14 males and 6 females, 25-57 years old). However, due to data unavailability, signals from subject 5 were not included, resulting in the use of data from 19 subjects. HD-sEMG signals were recorded using a Quattrocento (OT Bioelettronica) biomedical amplifier system, employing two 8×8 electrode grids, amounting to a total of 128 channels, with a 10 mm inter-electrode distance. As depicted in Fig. 1, one grid was positioned on the dorsal (outer forearm), and the other on the volar (inner forearm) of the upper forearm.

During data collection, each subject was instructed to perform each gesture five times consecutively, followed by a transition to the next one. The duration of each repetition was five seconds, followed by an equal-duration rest period. The sample rate is 2048Hz, and the Quattrocento device employed a hardware high-pass filter of 10Hz and a low-pass filter of 900Hz. The recording was performed differentially, with the signal for channel $i$ representing the difference between electrode $i+1$ and electrode $i$. The HD-sEMG data was segmented with a sliding time windows of 250ms with a 50% overlap.

## B. Experimental Setup

Our proposed method exclusively utilizes raw HD-sEMG data without any hand-crafted features. The construction of fully-connected muscle networks is based on the HD-sEMG series, where the number of nodes $N$ is set to 128, and each node is associated with the number of time samples in a sliding window, totaling 512. The determination of edges depends on the k-NN strategy to retain the strongest correlations, which is introduced in Session II.A. In this study, we evaluate the decoding performance for different values of $k$ in the k-NN approach, specifically a range of [2, 6] with an interval of 1. Ultimately, we set $k$ =2 to achieve a high prediction accuracy while maintaining sufficient sparsity in the muscle network graph.

The STGCN-GR model starts by utilizing a temporal convolutional network architecture to extract a lower-dimensional embedding representation from the HD-sEMG time series in each node. This process involves a layer of 1D convolutions followed by a GLU. The temporal convolutional kernel size $K_t$ is set to 5. Next, the graph convolutional network is applied, using a single layer of graph convolution. A ReLU function is used as the activation after each spatial convolution. After this step, layer normalization is performed over the node's features.

The model is trained for 100 epochs. The batch size is 64. Dropout is 0.5. The stochastic gradient descent with Adam optimizer is employed with an initial learning rate of 0.001 decay the learning rate by 0.05. Learning stops early after the validation loss plateaued (patience = 30). During training, the cross-entropy loss function is utilized as the optimization metric.

To evaluate the performance of our method, we adopt a five-fold cross-validation approach and report the mean and standard deviation (std) of validation accuracy as performance metrics. The experiments are implemented using PyTorch 2.0.0 and trained using the NVIDIA Tesla V100-PCIE GPU for computational acceleration.

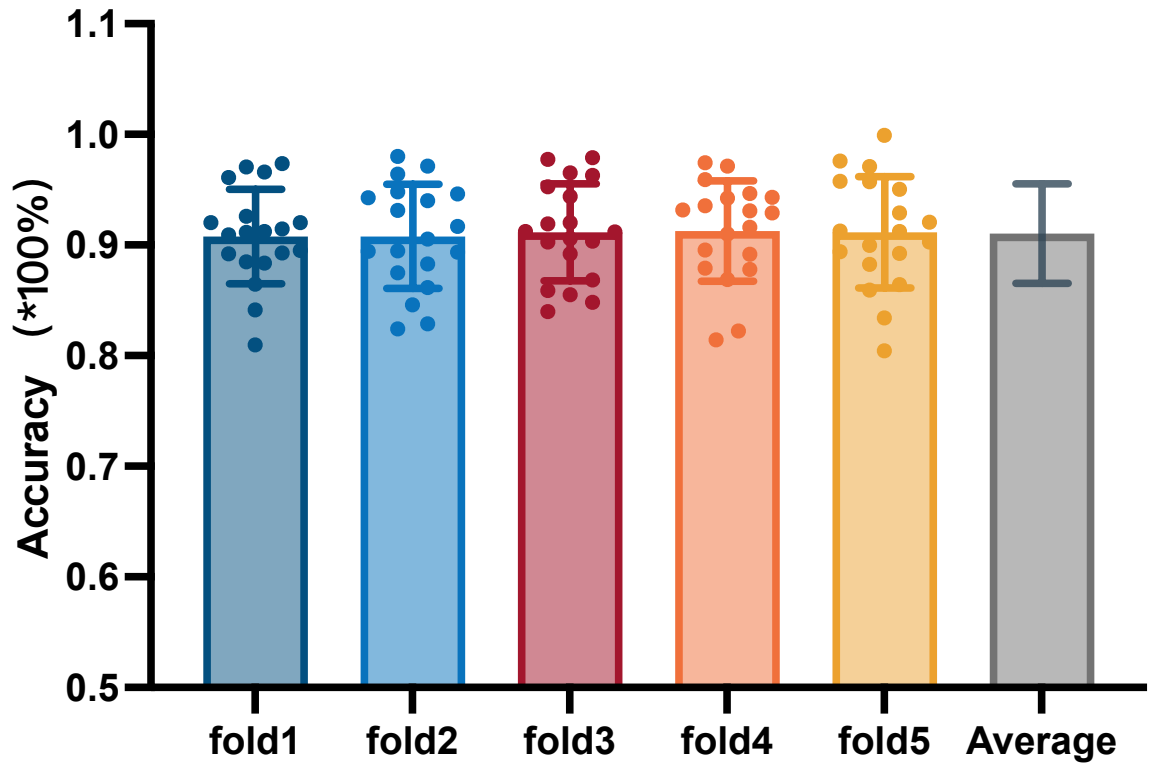

Fig. 3. The average classification accuracy of five-fold cross validation among 19 subjects.

TABLE I. COMPARISONS OF THE AVERAGE ACCURACIES (%) OF FIVE-FOLD CROSS CALIDATION AMONG THE VARIOUS METHOS. THE BEST VALUE ARE HIGHLIGHTED IN BOLD

| *Work* | *Model* | *Window length(ms)* | *Accuracy (%)* |
| --- | --- | --- | --- |
| T. Sun *et. al* [19] | Deep Heterogeneous Dilation of LSTM | 200 | 83.3 |
| N. Malešević *et. al* [20] | ViT-HGR | 31.25 | 84.62 |
| N. Malešević *et. al* [21] | HGR-Macro Model | 250 | 89.34 |
| **Proposed mothod** | STGCN-GR | 250 | **91.07** |

## IV. RESULT AND DISCUSSION

### A. Graphs of Muscle Networks

Fig. 2 presents intriguing graphs showcasing full-connected muscle networks constructed from the 128 electrode signals, consisting of 64 forearm extensor and 64 forearm flexor signals, for one subject. While the dataset comprises 65 gestures, to accommodate limited space, we illustrate only eight representative gestures in the graph. According to the electrode location, we partitioned the 128 nodes into eight distinct regions (R1-R8) and color-coded them accordingly. Nodes of the same color have similar geographic locations. Thicker edges within the graphs indicate stronger correlations between nodes.

The muscle network diagrams exhibit remarkable variability across the diverse gestures. For instance, gesture Palma grasp exhibits a conspicuous interdependence between nodes in regions R4 and R5, suggesting a coordinated relationship between these specific muscle groups during the execution of this gesture. In gesture Finger stretch, the extensor region displays a robust correlation among nodes R2, R3, and R4, indicating a close functional association between these muscles. On the other hand, the correlation within the flexor region appears to be relatively weaker, implying a different coordination pattern for this specific gesture. In addition, we can find that some nodes have strong interdependencies with each other, even though they are geographically distant. The observed dependencies and correlations provide valuable insights into the underlying muscle coordination during specific movements, shedding light on the complex interplay between muscle groups during hand gestures.

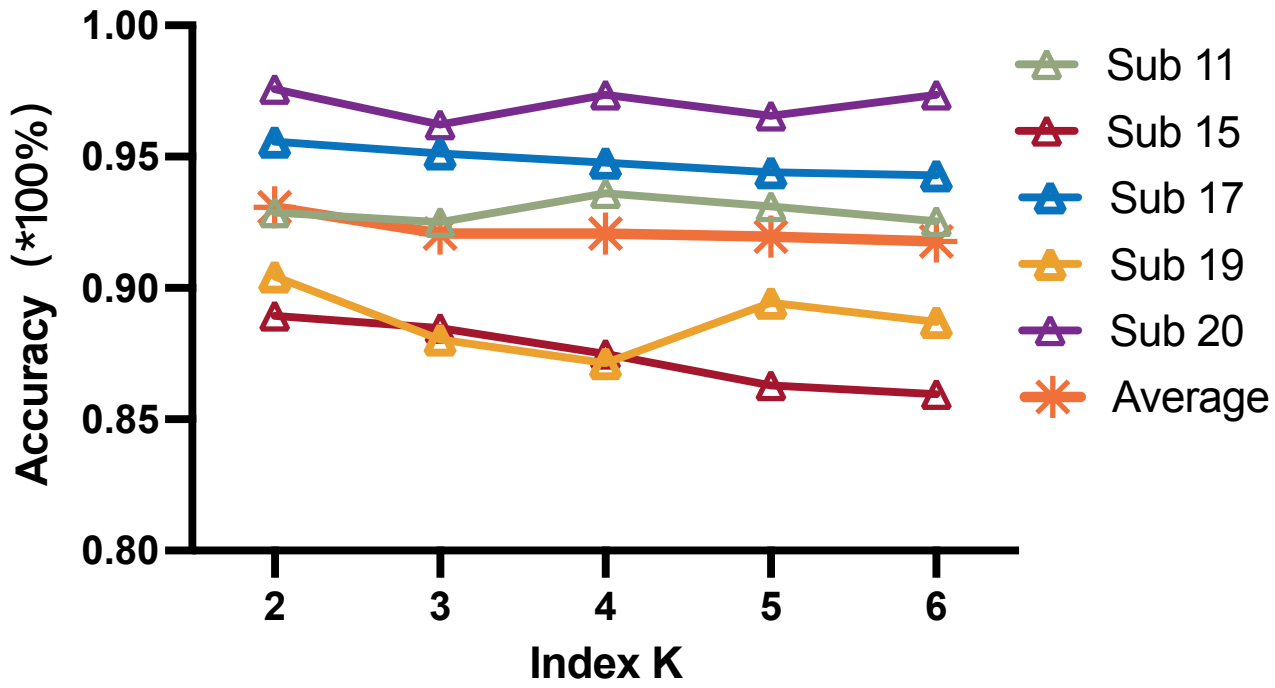

Fig. 4. The corresponding model accuracy for parameter k over a range from 2 to 6 with an interval of 1.

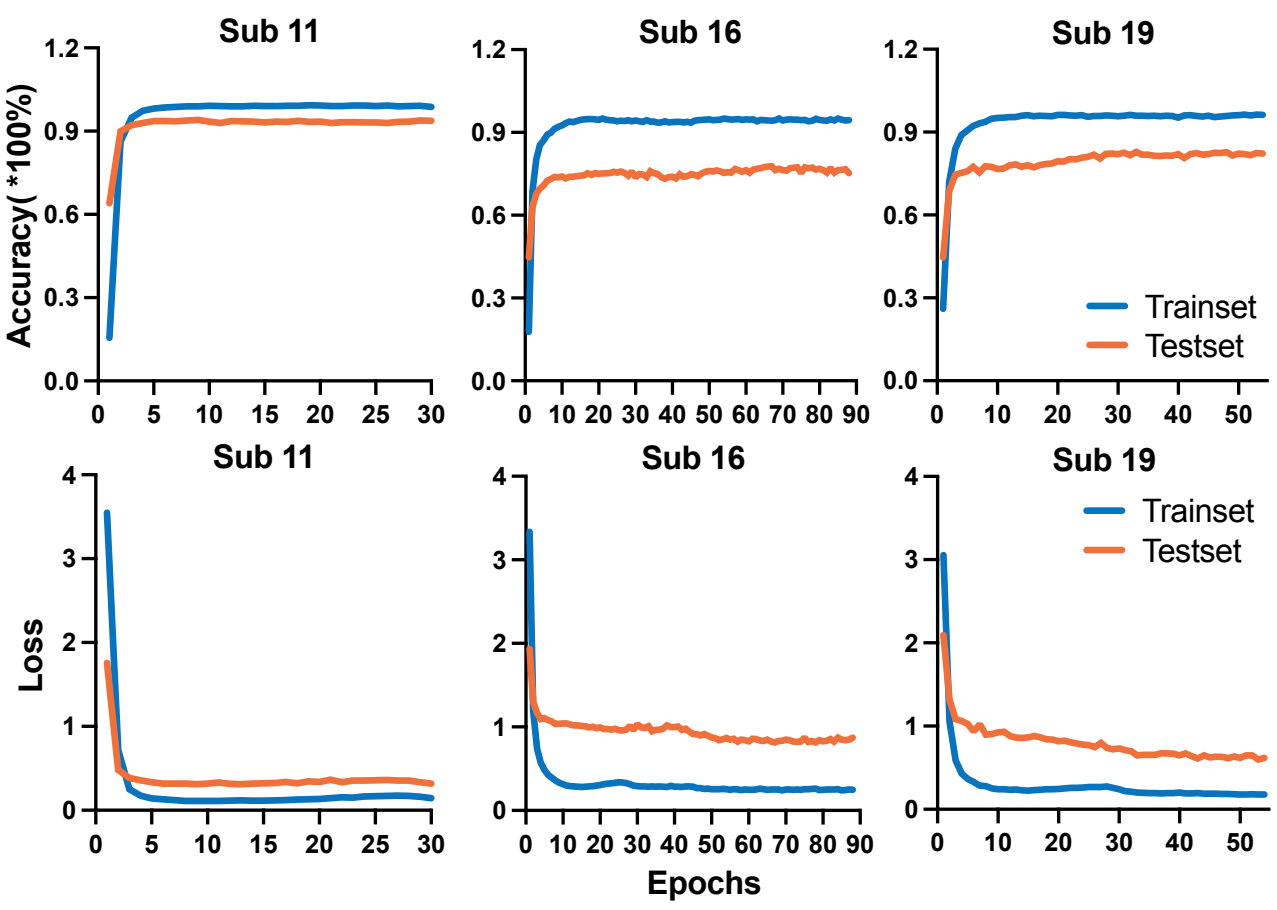

Fig. 5. The training and testing accuracy and loss of the model on different datasets as iterations progress.

### B. Gesture Decoding Performance and Comparison

To evaluate the model performance on the dataset, we adopt a rigorous five-fold cross-validation approach. During each fold, one repetition is held out for testing, and the remaining four repetitions are utilized for training the model. The average results of the cross-validation for the 19 subjects are visually presented in Fig. 3. Notably, the achieved accuracies for each fold are 91.43 ± 3.90%, 90.78 ± 4.58%, 91.15 ± 4.28%, 91.26 ± 4.40%, 91.15 ± 4.90%, respectively. The overall average accuracy among subjects across all folds is an impressive 91.07 ± 4.13%.

Furthermore, we conduct an in-depth comparison with other deep learning-based methods that utilized the same dataset [19-21]. These works also employed five-fold cross-validation during their evaluations. To ensure a robust comparison, we directly quote the results reported in the literature. The experimental outcomes are thoroughly analyzed and presented in Table I. The findings are compelling as they demonstrate that the proposed STGCN-GR method consistently outperforms other methods, achieving a higher classification accuracy. These results serve as compelling evidence of the effectiveness and superiority of STGCN-GR for gesture decoding tasks.

### C. Parameter Analysis

Within this section, we perform experiments on datasets involving five subjects to assess the impact of the $k$ parameter in the k-NN graph. We systematically vary the value of $k$ over a range from 2 to 6 with an interval of 1. Fig. 4 displays the corresponding model performance for different $k$. For the represented subjects, results demonstrate that the optimal accuracy is attained when $k$ is set to 2, resulting in a classification accuracy of 93.07% across five subjects.

A noteworthy observation is that the classification accuracy does not exhibit significant fluctuations with different values of $k$ for each subject. This finding suggests that the proposed model displays minimal sensitivity to changes in the number of edge connections within the muscle network graph. Understanding and acknowledging such variations can be crucial in the design and optimization of gesture recognition systems. This insight ensures the model's robustness and reliability in various applications involving HD-sEMG data.

### D. Model Converge Analysis

Fig. 5 presents a graphical representation of the model's training and testing accuracy and loss across three subjects during a representative run. The results reveal a notable trend where both training and testing accuracy show significant improvements during the initial epochs, followed by a gradual convergence to a stable level. In contrast, the corresponding loss experiences a rapid decrease at the onset of training, eventually leveling off with only marginal changes afterward.

An important finding from this analysis is that both the accuracy and loss of the model have reached a state of stability within the first 15 epochs in a smooth manner. This robust and smooth convergence within a relatively short training period showcases the effectiveness of the model's learning process. It indicates that the model has effectively captured the underlying patterns in the data and can make reliable predictions without overfitting.

## V. Conclusions

This paper presents STGCN-GR, a novel gesture recognition method for HD-sEMG-based HMIs. By combining graph convolution and temporal convolution modules, the proposed model effectively leverages spatial topological information and captures temporal dependencies inherent in the graph-based HD-sEMG presentation. Assessments are conducted on a diverse dataset of HD-sEMG signals from 19 able-bodied subjects performing 65 different gestures, demonstrating the model's adaptability to varying spatio-temporal features. The experimental results highlight the superiority of STGCN-GR over state-of-the-art deep learning methods for the same dataset, particularly with a sliding window size of 250ms, surpassing the real-time implementation requirement of 300ms. The STGCN-GR achieves a satisfactory classification accuracy and demonstrates notable generality, showcasing its potential for practical applications in real-world scenarios involving gesture recognition and HMI systems.